\documentclass[aps,prr,reprint, amsmath, amssymb,superscriptaddress]{revtex4-1}

\pdfoutput=1
\usepackage[utf8]{inputenc}
\usepackage{bm}
\usepackage[english]{babel}
\usepackage[T1]{fontenc}
\usepackage{mathrsfs}
\usepackage[retainorgcmds]{IEEEtrantools}
\usepackage{amsmath}
\usepackage{amssymb}
\usepackage{color}
\usepackage{amsfonts}
\usepackage{times,txfonts}
\usepackage{nicefrac}
\usepackage[colorlinks=true,linkcolor=blue,urlcolor=blue,citecolor=blue,pdfusetitle]{hyperref}

\usepackage{float}
\usepackage{epsfig} 
\usepackage{subfigure}

\usepackage{physics}
\usepackage{amssymb}
\usepackage[amssymb]{SIunits}

\usepackage{lipsum}
\usepackage{enumerate}

\newcommand{\dg}[1]{\ensuremath{#1^{\dagger}}}

\usepackage[dvipsnames]{xcolor}

\begin{document}

\title{Dephasing enhanced transport in boundary-driven quasiperiodic chains}

\author{Artur M. Lacerda}
\affiliation{Instituto de Física, Universidade de São Paulo, CEP 05314-970, São Paulo, São Paulo, Brazil}
\affiliation{Department of Physics, Trinity College Dublin, Dublin 2, Ireland}
\author{John Goold}
\affiliation{Department of Physics, Trinity College Dublin, Dublin 2, Ireland}
\author{Gabriel T. Landi}
\affiliation{Instituto de Física, Universidade de São Paulo, CEP 05314-970, São Paulo, São Paulo, Brazil}

\begin{abstract}
We study dephasing-enhanced transport in boundary-driven quasi-periodic systems. Specifically we consider dephasing modelled by current preserving Lindblad dissipators acting on the non-interacting Aubry-André-Harper (AAH) and Fibonacci bulk systems. The former is known to undergo a critical localization transition with a suppression of ballistic transport above a critical value of the potential. At the critical point, the presence of non-ergodic extended states yields anomalous sub-diffusion. The Fibonacci model, on the other hand, yields anomalous transport with a continuously varying exponent depending on the potential strength. By computing the covariance matrix in the non-equilibrium steady-state, we show that sufficiently strong dephasing always renders the transport diffusive.  
The interplay between dephasing and quasi-periodicity gives rise to a maximum of the diffusion coefficient for finite dephasing, which suggests the combination of quasi-periodic geometries and dephasing can be used to control noise-enhanced transport. 


\end{abstract}

\maketitle

\section{Introduction}

Non-equilibrium systems are characterized by the existence of macroscopic currents of energy or matter~\cite{de2013non}. Understanding these transport properties has, for more than a century, been a major field of research in physics. One of the fundamental issues is to ascertain the necessary ingredients in a model to induce a certain transport regime. Low dimensional systems, in particular, have received significant attention both in the classical~\cite{lepri2003thermal,dhar2008heat} and quantum regimes~\cite{bertini2021finite}. For instance, chains of harmonic oscillators present ballistic transport~\cite{Rieder1967}, which is fundamentally different from the diffusive behavior obtained in Fourier's law of heat conduction~\cite{Tritt2004}.
However, not all anharmonicities lead to diffusivity~\cite{Aoki2000a,Aoki2006}. In addition to being fundamental, being able to understand and control transport offers opportunities for potential device applications. Transport in low-dimensional devices, for instance, can be manipulated for steady state thermal machines~\cite{Benenti2017a} and nano-scale heat engineering~\cite{Terraneo2002,Li2004a,Li2005,Li2012}. 

For low dimensional quantum many-body systems, the technique of boundary-driving has been used widely to extract high temperature transport properties in non-equilibrium steady-states (NESSs)~\cite{PhysRevLett.102.207207,prosen2009matrix,vznidarivc2010exact,prosen2011exact,vznidarivc2011spin,mendoza2013dephasing,mendoza2013heat,karevski2013exact,landi2015open}. Boundary-driving is an open systems technique where local Lindblad dissipators at the edges of the chain induce a gradient in spin and or energy. Transport coefficients can then be extracted from finite size scaling of the currents~\cite{bertini2021finite,Landi2021}. Although the technique is limited to high temperature physics it has a distinct advantage in so much as it can be extended to models with integrability breaking perturbations~\cite{PhysRevB.98.235128,PhysRevLett.125.180605} by means of tensor networks. This technique has been instrumental in shedding light on the transport properties of the ergodic phase of interacting disordered models~\cite{Znidaric2016,Znidaric2016b,juanjose,schulz2020phenomenology} and systems with magnetic impurities~\cite{PhysRevB.98.235128,PhysRevLett.125.180605}.

Another class of models where boundary driving has been applied successfully is quasi-periodic chains~\cite{Varma2017,Varma2019,vznidarivc2018interaction}. A paradigmatic example is the Aubry-André-Harper (AAH) model~\cite{Harper1955, Aubry1980}, which is known to undergo a localization transition when the potential strength is increased. The model is readily simulated in ultra-cold atomic physics experiments, using bichromatic optical lattices, {\color{black} and on photonic lattices.~\cite{Roati2008, Lahini2008}. It has also been used} to  explore the many-body localised phase~\cite{schreiber2015observation}.   In the AAH, at fixed tunneling rate and below a critical value of the potential strength, all the energy eigenstates are delocalized, while above this value the entire spectrum is localized. This transition is clearly reflected in the non-equilibrium transport properties, with particle transport going from ballistic to exponentially suppressed. At criticality, the eigenstates are neither delocalized nor localized, and the transport is sub-diffusive~\cite{Purkayastha2017b, Varma2017}.
Generalizations of the AAH model exhibiting  mobility-edges have also been analyzed in the transport scenario~\cite{Purkayastha2017}. A closely related model is the Fibonacci model~\cite{Ostlund1983, Kohmoto1987, Abe1988, HIRAMOTO1992}. This model has a critical spectrum which has been studied in detail~\cite{Ostlund1983, Kohmoto1983, Abe1988, HIRAMOTO1992}. This unique spectrum gives rise to highly anomalous transport in the absence of interactions where the transport exponent varies continuously with the potential strength. In fact, the transport in the Fibonacci model can be tuned continuously, from ballistic to sub-diffusive.  While the quantum transport properties of quasi-periodic systems is well studied, recently studies have highlighted that they can be exploited for thermal engineering~\cite{PhysRevLett.123.020603,Chiaracane2020}.


In this work we are interested in how the anomalous transport, typically observed in quasi-periodic systems, is affected by dephasing noise from an environment. One phenomenological approach to model dephasing is by means means of self-consistent baths~\cite{Bolsterli1970}, also known as B\"uttiker probes~\cite{Buttiker1985}. Another approach, which we will exploit here, is via current-preserving Lindblad dissipators.
For any non-zero dephasing strengths, free tight-binding models typically become diffusive in the thermodynamic limit~\cite{vznidarivc2010exact, Asadian2013}. Motivated by these results, we study  the effects of dephasing in quasiperiodic systems. We numerically compute the covariance matrix in the NESS and  study the finite size scaling of the particle current. Both the AAH model and the Fibonacci model become diffuive in the presence of dephasing noise. Interestingly, though, in certain regimes this introduces a competition, where the dephasing leads to noise-enhanced transport. 
The paper is divided as follows. The model is described in Sec.~\ref{sec:model}
and the analysis of the transport properties of both models, with and without dephasing, are discussed in Sec.~\ref{sec:analysis}.
Our main results are presented in Sec.~\ref{section:dephasing_enhanced}, where we analyze the interplay between quasi-periodicity and dephasing. Conclusions are summarized in Sec.~\ref{sec:discussion}.

\section{\label{sec:model}The model}

\subsection{Boundary-driven \textit{XX} chain}

Consider a one-dimensional spin 1/2 (or free Fermion) system with $L$ sites, described by the \textit{XX} (tight-binding) Hamiltonian 
\begin{IEEEeqnarray}{rCl}
\label{hamiltonian_spin}
    H &=& -\sum_{i=1}^{L-1}\qty(\sigma_i^+\sigma_{i+1}^- + \sigma_i^-\sigma_{i+1}^+) + \frac{\lambda}{2}\sum_{i}^L V_i\sigma_i^z\\[0.2cm]
    \label{hamiltonian_fermions}
    &=& -\sum_{i=1}^{L-1}\qty(c_i^\dag c_{i+1} + c_{i+1}^\dag c_i) + \lambda\sum_{i}^L V_i c_i^\dag c_i,
\end{IEEEeqnarray}
which are equivalent via a Jordan-Wigner transformation $c_i = \prod_{j=1}^{i-1} (-\sigma_i^z) \sigma_i^-$. We consider two models, defined by two different choices of the onsite potential $V_i$, with strength $\lambda$ (Fig.~\ref{fig:no_deph}(a)). 
The first is the Aubry-André-Harper (AAH) model, in which the on-site potential is given by~\cite{Harper1955, Aubry1980}
\begin{equation}
    \label{eq:aubry_andre}
    V_i = 2\cos(2\pi g i + \theta),
\end{equation}
where $g = (1+\sqrt 5)/2$ is the golden ratio.
This model undergoes a localization transition at $\lambda_c = 1$.  
When $\lambda<1$, all  energy eigenstates are delocalized, and when $\lambda>1$ they are all localized. 
Similar results also follow when $g$ is any other Diophantine number \cite{Jitomirskaya1999}. The second system is the Fibonacci model, with a potential is defined by~\cite{Ostlund1983, Kohmoto1987, Abe1988, HIRAMOTO1992}
\begin{equation}
    \label{eq:fibonacci}
    V_i = \qty[\frac{i+1}{g^2}] - \qty[\frac{i}{g^2}],
\end{equation}
where $[x]$ is the integer part of $x$. This represents  the $i$th element of a binary sequence called Fibonacci word, which can be constructed by the recursive rule $S_n = S_{n-2} + S_{n-1}$, with $S_0=0$, $S_1=01$ and  ``+''  taken as concatenation. Successive application of this rule generates the words:
$$
    0 \rightarrow 01 \rightarrow 010 \rightarrow 01001 \rightarrow 01001010 \rightarrow \cdots
$$
Notice that each Fibonacci word is an extension of the previous one; $V_i$ is then the $i$th digit of any word with size greater or equal to $i$. Additionally, the length of each word is a Fibonacci number, by construction~\cite{Ostlund1983, Kohmoto1987, Abe1988, HIRAMOTO1992, Varma2019}. 

We consider both of these systems, driven out of equilibrium by boundary reservoirs and every site  subject to dephasing noise. The time evolution of the density matrix is described via the Gorini, Kossakowski and Sudarshan, Lindblad (GKSL) master equation \cite{Gorini1976, Lindblad1976},
\begin{equation}
    \label{eq:master_eq}
    \dv{\rho}{t}= -i\comm{H}{\rho} + D_1(\rho) + D_L(\rho) + \sum_{i=1}^L D^\text{deph}_i(\rho).
\end{equation}
The dissipators $D_i$ describe the action of the two driving baths at the boundaries~\cite{Landi2021}, and are given by
\footnote{ If one uses instead the dissipators in the spin formulation, $\gamma (1 - f_i)\mathcal D[\sigma^-_i] + \gamma f_i \mathcal D[\sigma^+_i]$, the current remains unchanged, even though the JW-transformation introduces a parity term in the last site.}
\begin{equation}
    \label{eq:boundary_dissipator}
    D_i(\rho) = \gamma (1 - f_i)\mathcal D[c_i] + \gamma f_i \mathcal D[c_i^\dagger] \qc i=1,L,
\end{equation}
where $\gamma$ is the coupling strength, $f_i$ is the Fermi-Dirac distribution of the bath and $\mathcal D$ is a Lindblad operator of the form
\begin{equation}
    \mathcal D[L] = L\rho \dg L - \frac12\acomm{\dg L L}{\rho}.
\end{equation}
Similarly, $D^\text{deph}_i$ in Eq.~\eqref{eq:master_eq} describes the dephasing on site $i$, and is given by 
\begin{IEEEeqnarray}{rCl}
    D^\text{deph}_i(\rho) &=& 
     \Gamma\mathcal D[c_i^\dag c_i] 
    \label{eq:dephasing_fermionic}
\end{IEEEeqnarray}
where $\Gamma$ is the dephasing strength. 


If the sites were uncoupled,  a bath of the form \eqref{eq:boundary_dissipator} would lead to the equilibrium state $\rho_\text{eq} = f\ketbra{0}{0} + (1-f)\ketbra{1}{1}$, 
where $\ev{\sigma^z}=2f-1$ and $\ev{c^\dagger c}=f$. Hence, the difference $\Delta f = f_L - f_1$ can be interpreted as either a magnetization or a population imbalance in the chain. As long as $\Delta f \neq 0$, the system will converge to non-equilibrium steady state (NESS) with a  non-zero magnetization (particle) current, given by 
\begin{equation}
    \label{eq:currrent}
    J_i = i\ev{c_{i+1}^\dag c_i - c_i^\dag c_{i+1}}.
\end{equation}
For the internal sites, $i = 2,\ldots, L-1$, these currents satisfy a continuity equation 
\begin{equation}
    \label{eq:continuity_eq}
    \dv{t}\ev{c_i^\dag c_i} = J_{i-1} - J_i \qc i=2,\dots, L-1,
\end{equation}
which is obtained directly from Eq.~\eqref{eq:master_eq}.
The sites at the boundaries are subject to additional currents $J_0 = \tr\big\{c_1^\dagger c_1 D_1(\rho)\big\}$ and $J_L = \tr\big\{c_L^\dagger c_L D_L(\rho)\big\}$.
Crucially, note that the dephasing dissipators do not affect the continuity equation~\eqref{eq:continuity_eq}. 
There is, therefore, no particle exchange with them. 

In the NESS, $\dd \ev{c_i^\dag c_i}/ \dd t = 0$. Hence, by Eq.~\eqref{eq:continuity_eq} the current becomes homogeneous throughout the  chain:
\begin{equation}
    J_1 = J_2 = \cdots = J_{L-1} \equiv J.
\end{equation}
We can thus unambiguously refer to the particle current simply as $J$. 
In the spin chain formulation, the particle current  naturally translates to a magnetization current,
\begin{equation}
    J_i = 2i\ev{\sigma^x_i\sigma^y_{i+1}-\sigma^y_i\sigma^x_{i+1}}.
\end{equation}
This definition can also be obtained by writing an explicit expression for $\dd \ev{\sigma^z_i}\!/\dd t$ and interpreting it as a continuity equation, similarly to Eq.~\eqref{eq:continuity_eq}.

\subsection{Steady-state equation for the Covariance Matrix}

The free fermion nature of this model allows us to focus only on the system's covariance matrix, defined as
\begin{equation}
    C_{ij} = \ev{c_j^\dag c_i},
\end{equation}
and from this the particle current can be extracted as
$J = 2 \Im C_{i,i+1}$.
The time evolution of $C$ can be obtained directly from Eq.~\eqref{eq:master_eq}  (see \cite{vznidarivc2010exact, Asadian2013, Malouf_2020} for details), and reads
\begin{equation}\label{C_eq}
    \dv{C}{t} = -(WC + C W^\dag) -\Gamma \Delta(C) + F
\end{equation}
where
\begin{align}
    &W_{ij} = -(\delta_{i+1, j} + \delta_{i, j+1}) + \lambda V_i \delta_{ij} -\frac{\gamma}{2}(\delta_{i,1}\delta_{j,1} +\delta_{i,L}\delta_{j,L}) ,\\[0.2cm]
    &F = \mathrm{diag}(\gamma f_1,0, ...,0, \gamma f_L),
\end{align}
and $\Delta(\cdot)$ is an operation that removes the diagonal of a matrix:
\begin{equation}
    \Delta (C) = C - \mathrm{diag}(C_{11}, C_{22}, ..., C_{LL}).
\end{equation}
In the NESS, $\dd C/\dd t = 0$, which leads to the matrix equation
\begin{equation}
    \label{eq:cm_ness}
    WC + C W^\dag + \Gamma \Delta(C) = F.
\end{equation}
When $\Gamma = 0$, this reduces to a {\color{black}Lyapunov} equation
\begin{equation}
    \label{eq:lyapunov_eq}
    WC + C\dg W= F,
\end{equation}
which can be efficiently solved numerically using the eigendecomposition method described in Ref.~\cite{Varma2017}. We have found that, at least in the parameter region explored, this method outperforms the standard solvers for Lyapunov equations. {\color{black} We solved Eq.~\eqref{eq:lyapunov_eq} for for sizes up to $L=1597$}.
When $\Gamma \neq 0$, Eq.~\eqref{eq:cm_ness} is still linear in $C$, but  not in Lyapunov-form. 
In this case we solve Eq.~\eqref{eq:cm_ness} using a standard solver for sparse linear systems. Since this system does not exhibit any special structure, besides its sparsity, the largest system size we were able to simulate is $L=987$, which is considerably smaller when compared to the $\Gamma = 0$ case.

\subsection{Classification of the transport regime}


In general, the current follows a power-law scaling with the system size:
\begin{equation}
    \label{eq:scaling_current}
    J \sim \frac{1}{L^\nu},
\end{equation}
where $\nu\geq 0$ is a transport coefficient. The transport is classified as ballistic if $\nu=0$, diffusive if $\nu=1$ and anomalous otherwise. Anomalous transport is further classified as superdiffusive if $0 < \nu < 1$ or subdiffusive if $\nu > 1$. The absence of transport can be seen as an extreme case of subdiffusion, where $\nu\to\infty$.
For non-interacting models the current is always proportional to the driving bias $\Delta f = f_L - f_1$~\cite{Asadian2013}, so we may in fact write $J \sim \Delta f / L^\nu$. Moreover, it depends only on the difference $\Delta f$, and not on the values $f_1$ and $f_L$ themselves. For this reason, we  henceforth fix $f_1=1$ and $f_L = 0$.
We also henceforth set $\gamma = 1$ in Eq.~\eqref{eq:boundary_dissipator}.

The coefficients $\nu$ are obtained  by computing the current for increasing values of $L$ and performing a linear regression of {\color{black}the form} $\log J = -\nu \log L + C$. The exact value of the coefficient $\nu$ may depend on the number-theoretic properties of the chosen family of sizes. And the quasiperiodicity usually makes the $L$ dependence somewhat noisy. To smooth this, we perform the regression using Fibonacci numbers for $L$~\cite{Varma2017}. Alternatively, we may also classify the transport properties through the system's finite-size conductivity $\kappa(L)$, which is defined from
\begin{equation}\label{conductivity}
    J = \kappa(L) \frac{\Delta f}{L}.
\end{equation}
Comparing this with Eq.~\eqref{eq:scaling_current}, we see that the conductivity must scale as $\kappa(L) \sim L^{1-\nu}$. 
It is therefore independent of $L$ only in the  diffusive case. 
For ballistic or {\color{black}superdiffusive} transport, it diverges when $L\to\infty$, whereas for subdiffusive transport it vanishes in this limit.

\section{\label{sec:analysis}Transport properties}

\subsection{Zero dephasing}

\begin{figure}
    \centering
    \includegraphics[width=0.48\columnwidth]{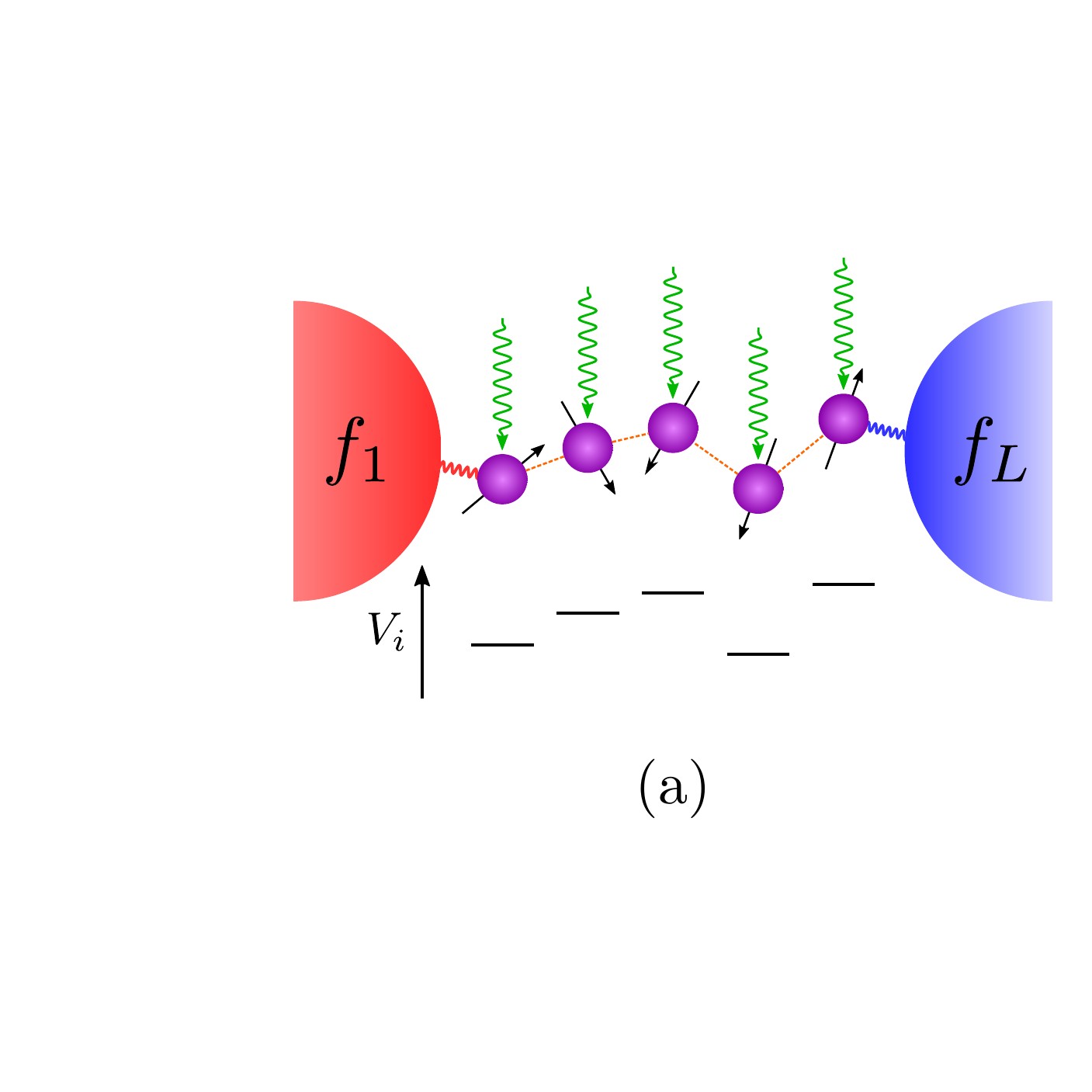}\quad
    \includegraphics[width=0.48\columnwidth]{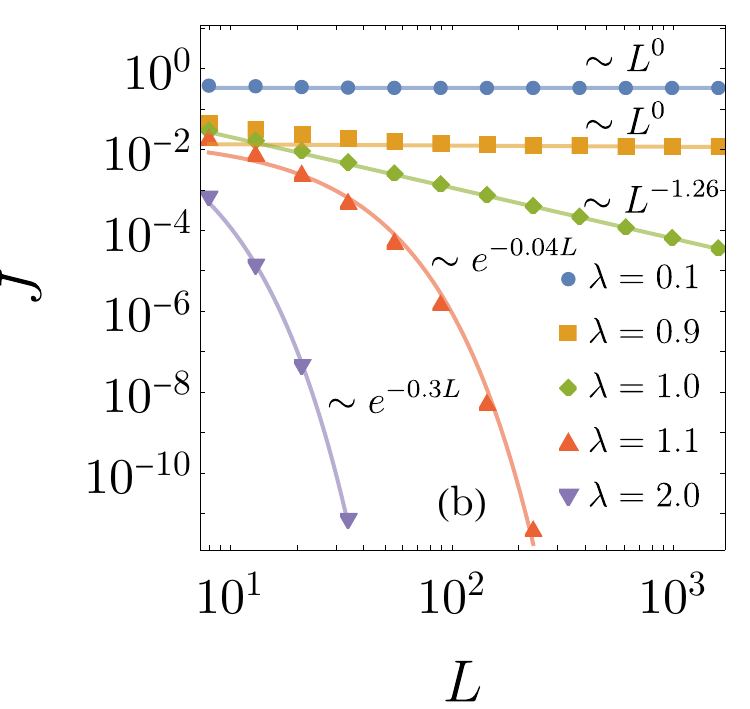}\\[0.2cm]
    \includegraphics[width=0.48\columnwidth]{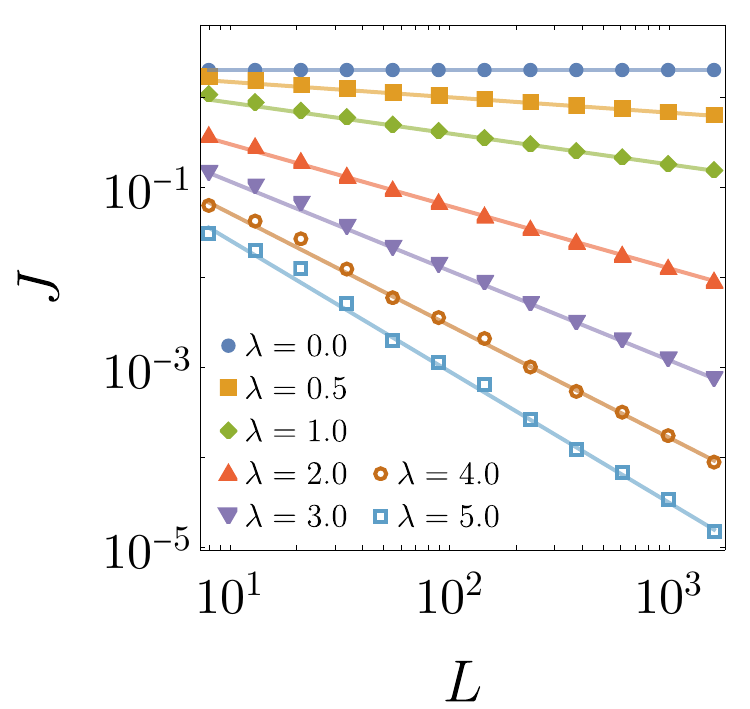}\quad
    \includegraphics[width=0.48\columnwidth]{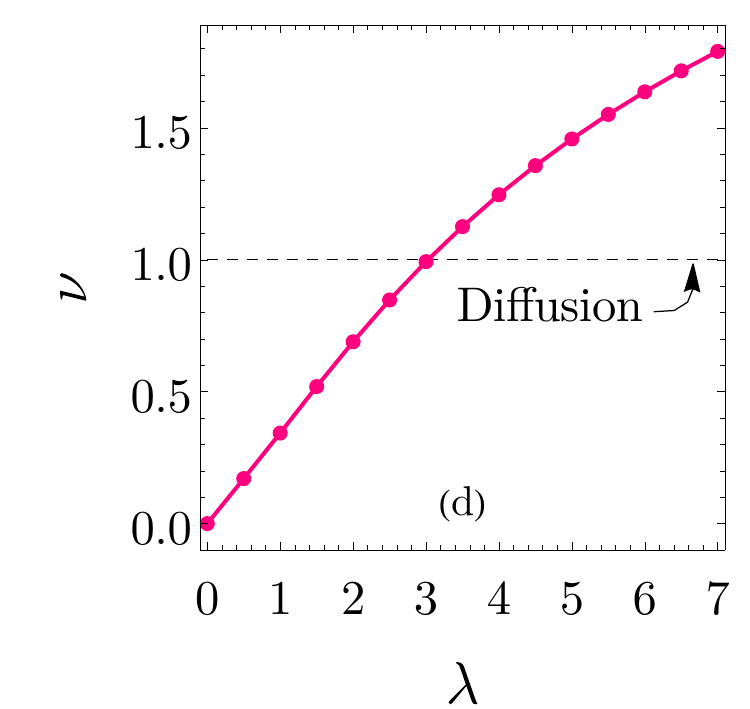}
    \caption{(a) Schematic representation of the boundary-driven quasiperiodic chains studied in this paper [Eq.~\eqref{hamiltonian_spin}]. (b)-(c) Summary of transport properties of the AAH and Fibonacci models, in the absence of dephasing ($\Gamma=0$). {\color{black}The solid lines indicate the scaling laws obtained by linear regression, using all the data points in the localized case and the last five points elsewhere.}
    (b) $J$~vs.~$L$ for the AAH model for different values of $\lambda$. 
    (c) Same, but for the Fibonacci model. 
    (d) Transport exponent $\nu$ [Eq.~\eqref{eq:scaling_current}] as a function of $\lambda$ for the Fibonacci model.
    }
    \label{fig:no_deph}
\end{figure}

Fig.~\ref{fig:no_deph} provides a summary of the transport properties without dephasing ($\Gamma=0$).
Fig.~\ref{fig:no_deph}(b) focuses on the  AAH model, for different disorder strengths $\lambda$.
All results are already averaged over 100 values of the phase $\theta$ [Eq.~\eqref{eq:aubry_andre}] to reduce fluctuations. 
The localization transition at $\lambda = 1$ is clearly reflected: 
For $\lambda <1$ the transport is ballistic, while for $\lambda > 1$ it decays exponentially (insulating). 
At $\lambda=1$ the transport is subdiffusive, with $\nu = 1.26$.
This  is  close to the value of 1.27 reported in \cite{Varma2017}. {\color{black} This discrepancy is likely due to the fact that in \cite{Varma2017} the authors computed the current up to larger system size, and averaged the results over a larger number of samples with different phases.}




Fig.~\ref{fig:no_deph}(c) shows the scaling for the Fibonacci model. 
As $\lambda$ increases, the slope of the curves become gradually more negative, causing the system to change continuously from ballistic (when $\lambda = 0$) to localized (when $\lambda \to \infty$). 
This is more clearly seen in Fig.~\ref{fig:no_deph}(d), which summarizes the dependence of  $\nu$ on $\lambda$, showing that the transport can be tuned to any regime. The diffusive point ($\nu = 1$) occurs  around $\lambda\approx3$.

\subsection{Non-zero dephasing}
\label{sec:transport_dephasing}

\begin{figure}
    \centering
    \includegraphics[width=0.48\columnwidth]{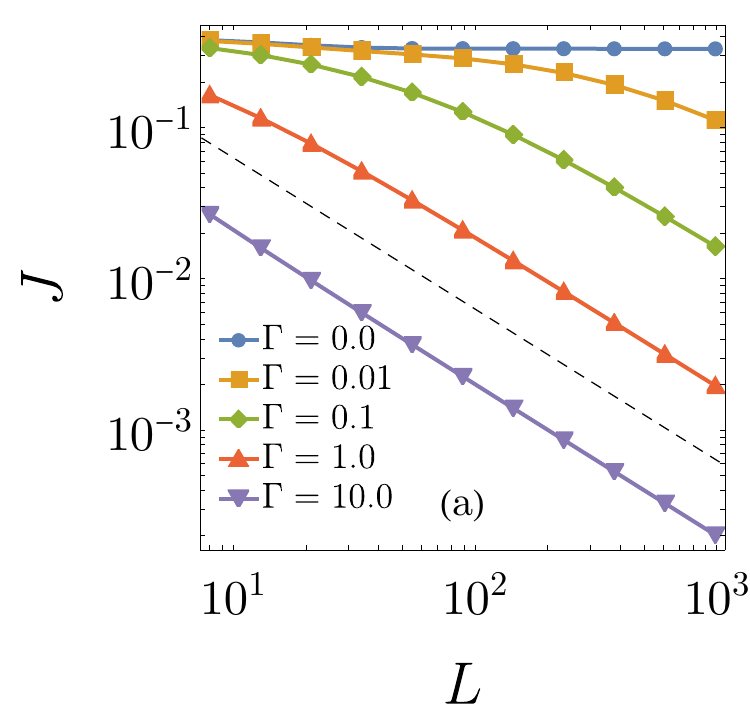}\quad
    \includegraphics[width=0.48\columnwidth]{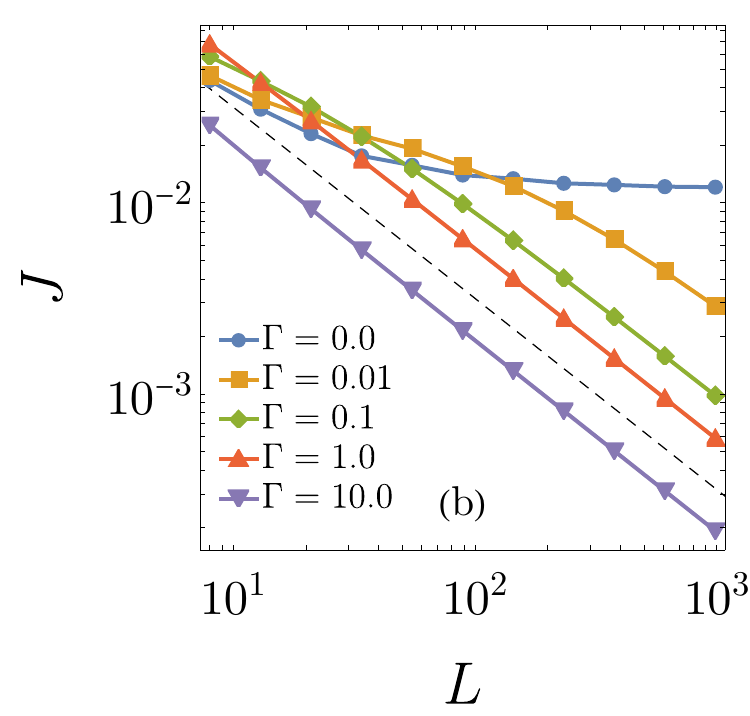}\\[0.2cm]
    \includegraphics[width=0.48\columnwidth]{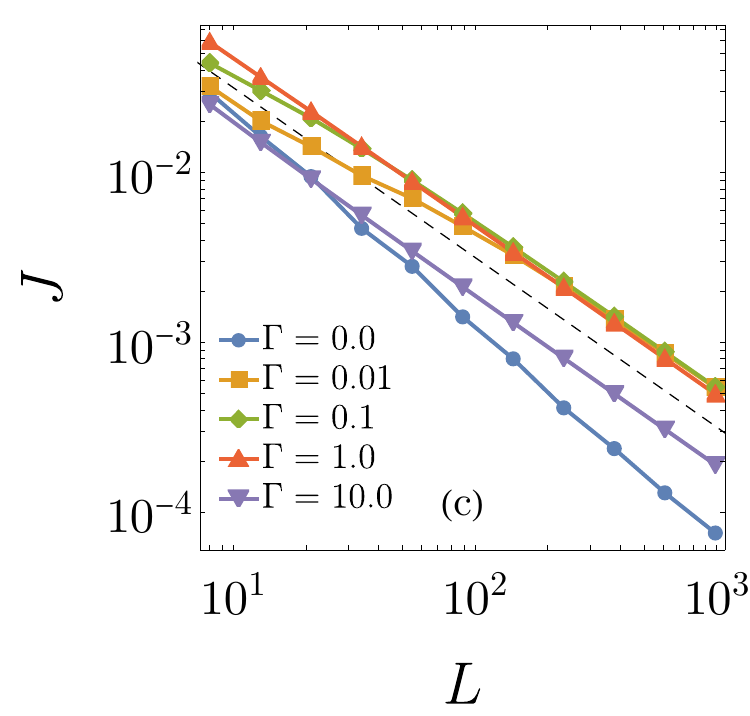}\quad
    \includegraphics[width=0.48\columnwidth]{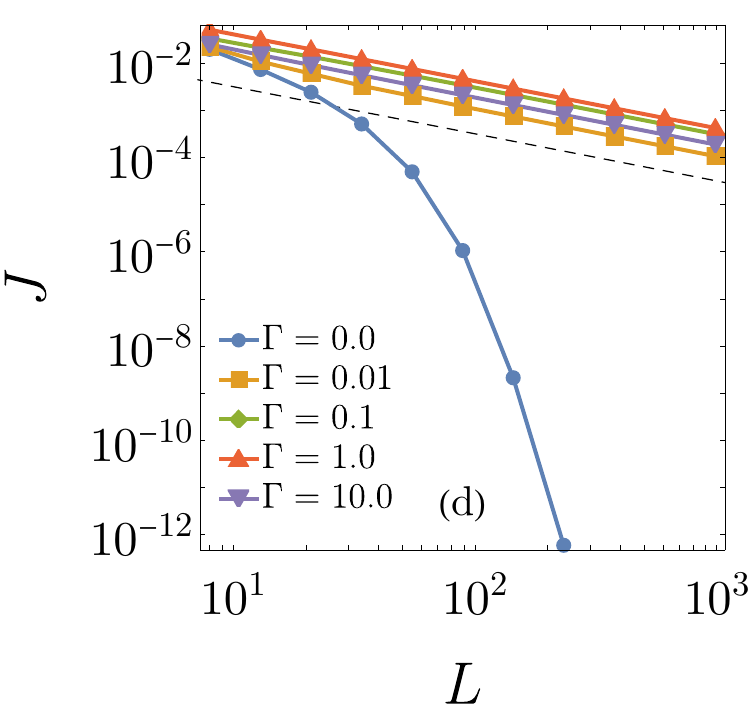}
    \caption{$J$~vs.~$L$ for the AAH model with different dephasing strengths $\Gamma$.  (a) $\lambda=0.1$; (b) $\lambda=0.9$; (c) $\lambda=1.0$; (d) $\lambda=1.1$. 
    The dashed line is a visual guide for the diffusive behavior, $J \propto L^{-1}$. 
    All results are averaged for 100 values of $\theta$, evenly spaced in between 0 and $\pi$. 
    The sizes $L$ are chosen as Fibonacci numbers, to reduce fluctuations.
    Other parameters: $\gamma=1$, $f_1=1$ and $f_L=0$. 
    }
    \label{fig:aah_deph}
\end{figure}

\begin{figure}
    \centering
    \includegraphics[width=0.48\columnwidth]{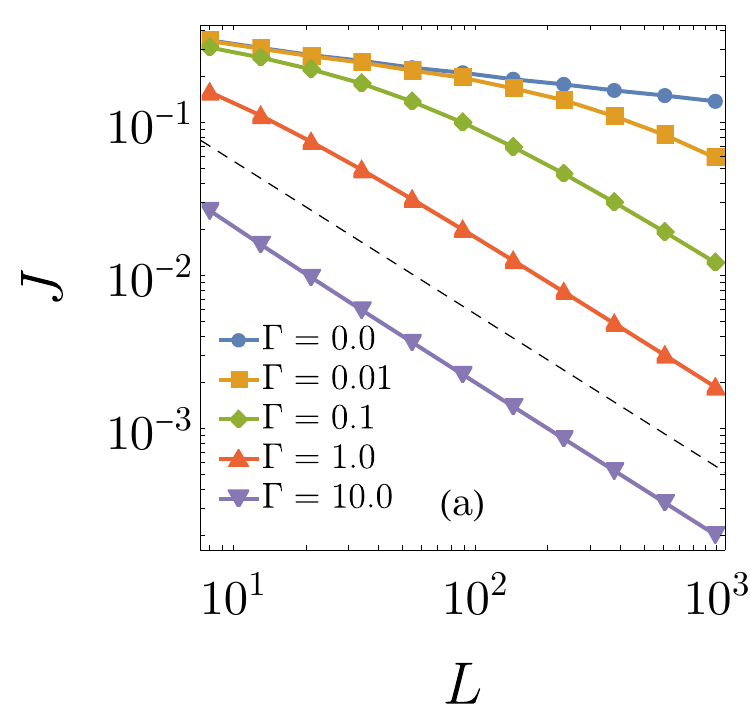}\quad
    \includegraphics[width=0.48\columnwidth]{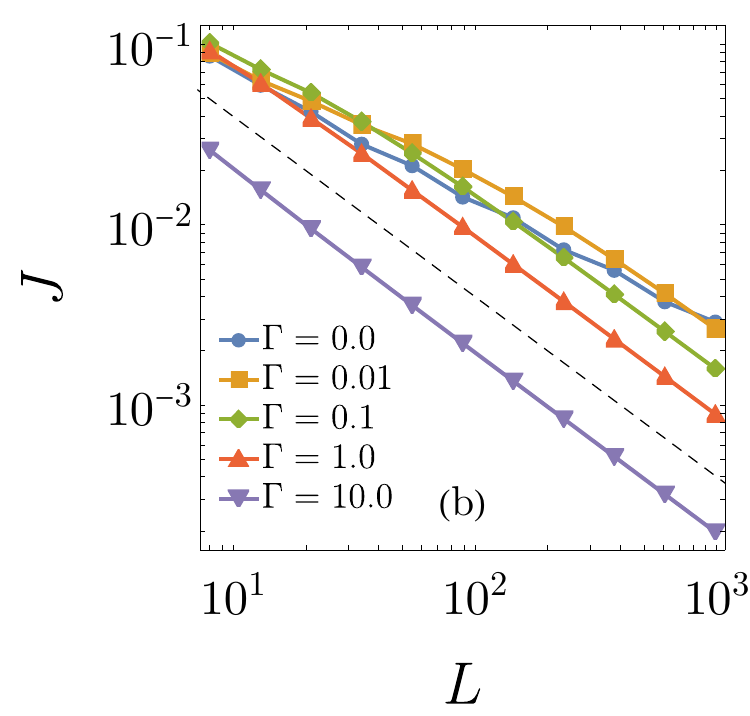}\\[0.2cm]
    \includegraphics[width=0.48\columnwidth]{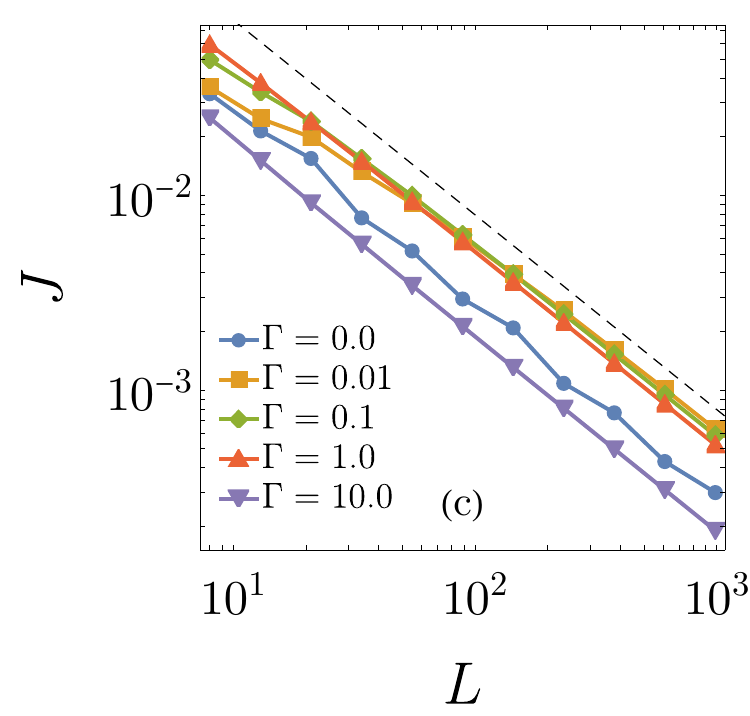}\quad
    \includegraphics[width=0.48\columnwidth]{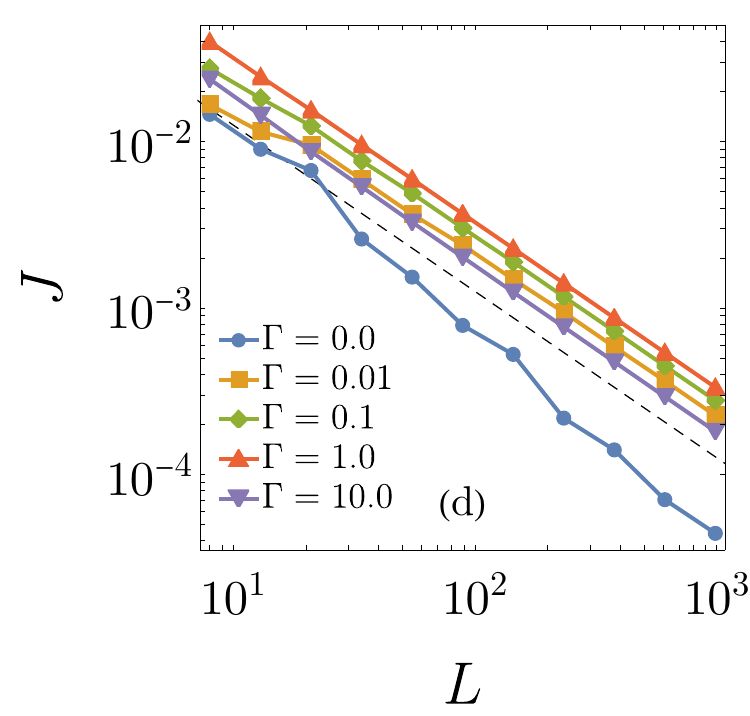}
    \caption{Similar to Fig.~\ref{fig:aah_deph}, but for the Fibonacci model. 
    (a) $\lambda=0.5$; (b) $\lambda=1.0$; (c) $\lambda=2.0$; (d) $\lambda=4.0$. 
    }
    \label{fig:fib_deph}
\end{figure}

Next we examine the effects of the addition of bulk dephasing. 
The properties of the AAH model are summarized in  Fig.~\ref{fig:aah_deph}, and the Fibonacci in Fig.~\ref{fig:fib_deph}. 
In both cases, dephasing always leads to diffusion for sufficiently large $L$, even for small values of $\Gamma$. 
This agrees with results from Ref.~\cite{Znidaric2013}, which studied disordered tight-binding chains. 
Figs.~\ref{fig:aah_deph}(c) and~\ref{fig:fib_deph}(d), in particular, illustrate scenarios where the bare transport ($\Gamma =0$) would be subdiffusive, but dephasing forces it to become diffusive. 
This indicates that dephasing may be used to generate enhanced transport, which will be discussed further in section \ref{section:dephasing_enhanced}. For any finite $\Gamma$, the dephasing will always render the transport diffusive for sufficiently large $L$. 
But the typical value of $L$ at which this takes place varies significantly in one regime or another (compare, e.g., Figs.~\ref{fig:aah_deph}(a) and (d)).
Ref.~\cite{Znidaric2016b} introduced  a characteristic length  $L_\Gamma$ for the dephasing effect to become important, which reads
\begin{equation}
    \label{eq:dephasing_length}
    L_\Gamma \sim 
    \Gamma^{-1/(1+\nu)}.
\end{equation}
{\color{black}This can indeed qualitatively describe some of the behavior in Figs.~\ref{fig:aah_deph} and~\ref{fig:fib_deph}. In Fig.~\ref{fig:aah_deph}(a), for instance, where $\nu \approx 0$, $L_\Gamma$ is large for small $\Gamma$. In contrast, in Fig.~\ref{fig:aah_deph}(d), where $\nu \gg 1$, the diffusive scaling sets in even for the smallest length scales.}

\section{Dephasing-enhanced transport}
\label{section:dephasing_enhanced}

After these preliminaries, we finally turn to the main result of this paper. Namely, that the combination of quasiperiodicity and dephasing can lead to the phenomenon of noise-enhanced transport. 
As discussed in Sec.~\ref{sec:transport_dephasing}, the addition of dephasing in both models always leads to diffusion, for any $\Gamma > 0$. 
However, when $\Gamma$ and $\lambda $ are both small, the original Hamiltonian should still play an important role when  $ L < L_\Gamma$ [Eq.~\eqref{eq:dephasing_length}]. 
A particularly convenient quantity for describing this interplay is the conductivity $\kappa$, defined in Eq.~\eqref{conductivity}.
Following~\cite{Znidaric2016b}, one expects that the existence of $L_\Gamma$
should cause $\kappa$ to present a  piecewise behavior with $L$:
\begin{equation}
    \label{eq:kappa_piecewise}
    \kappa(\Gamma, L) =
    \begin{cases} 
      L^{1-\nu} & L\leq L_\Gamma \\
      \kappa_{\text{deph}}(\Gamma) & L > L_\Gamma 
   \end{cases}
\end{equation}
Below $L_\Gamma$, it will in general depend on $L$, with coefficient $\nu$ dictated by the original transport properties of the system. 
But above $L_\Gamma$, the dephasing-induced diffusion will start to take place, so the conductivity must become a constant $\kappa_{\text{deph}}(\Gamma)$, independent of $L$ [as is characteristic of diffusive behavior]. 
The expression for $\kappa_{\text{deph}}$ can be obtained by imposing continuity on $L = L_\Gamma$, which results in
\begin{equation}
    \label{eq:kappa_deph_small_gamma}
    \kappa_{\text{deph}}(\Gamma) \sim L_\Gamma^{1-\nu} \sim \Gamma^{(\nu-1)/(\nu+1)} .
\end{equation}
These results, we emphasize, hold only for $\Gamma$ small. 

Conversely, when $\Gamma$ is much larger than the onsite potential $\lambda$, the effects of dephasing should be dramatic. 
The conductivity in this case can be obtained by simply setting $\lambda = 0$ in  Eq.~\eqref{eq:cm_ness}, which leads to \begin{equation}
    \label{eq:kappa_deph_large_gamma}
    \kappa_\text{deph}(\Gamma) \sim \frac{1}{\Gamma}.
\end{equation}
Notice that, for ballistic transport ($\nu=0$) the scalings in~\eqref{eq:kappa_deph_small_gamma} and~\eqref{eq:kappa_deph_large_gamma} coincide. 

\begin{figure}[h!]
    \centering
    \includegraphics[width=0.45\textwidth]{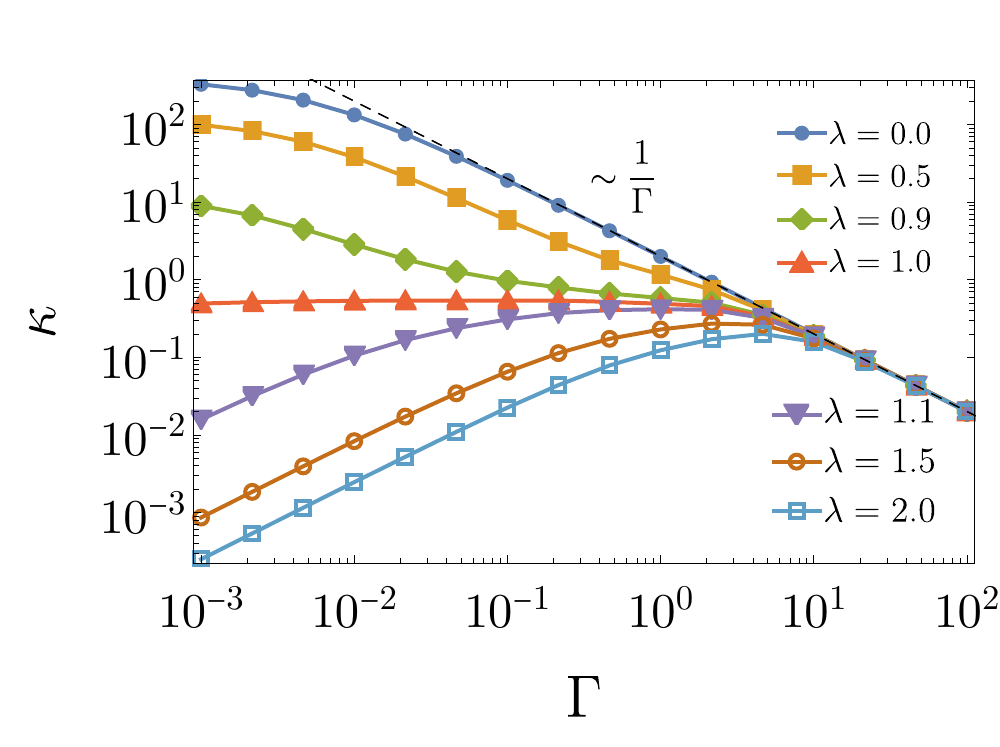}
    \caption{
    $\kappa$~vs.~$\Gamma$ for the AAH model, for  different values of $\lambda$, with fixed $L=987$. Other parameters are as in Fig.~\ref{fig:aah_deph}. 
    }
    \label{fig:dephasing_enhanced_aah}
\end{figure}
\begin{figure}[h!]
    \centering
    \includegraphics[width=0.4\textwidth]{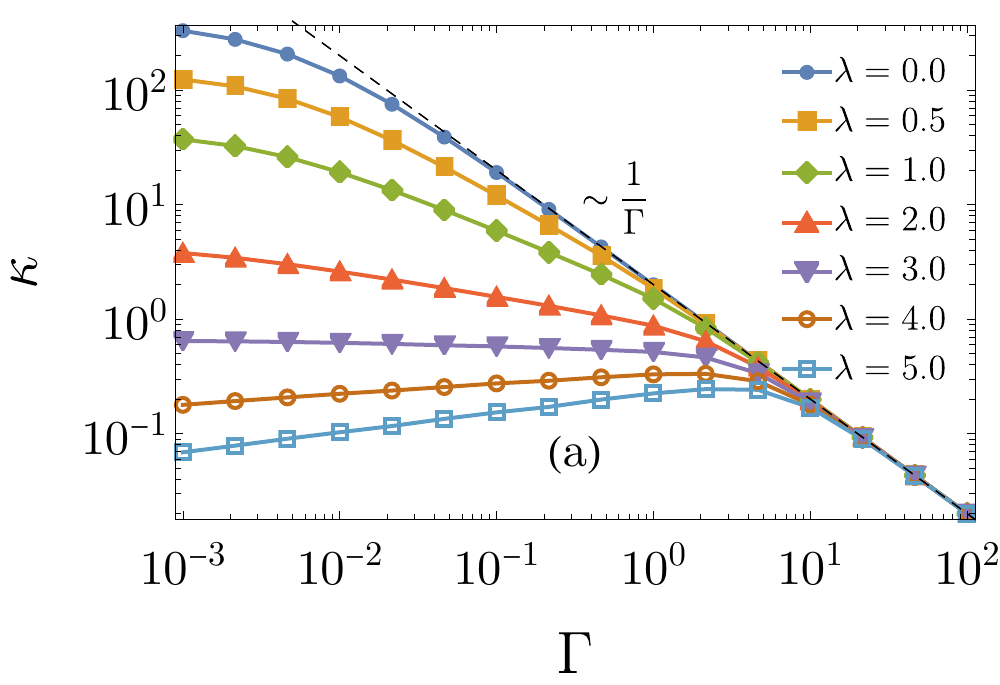}
    \includegraphics[width=0.4\textwidth]{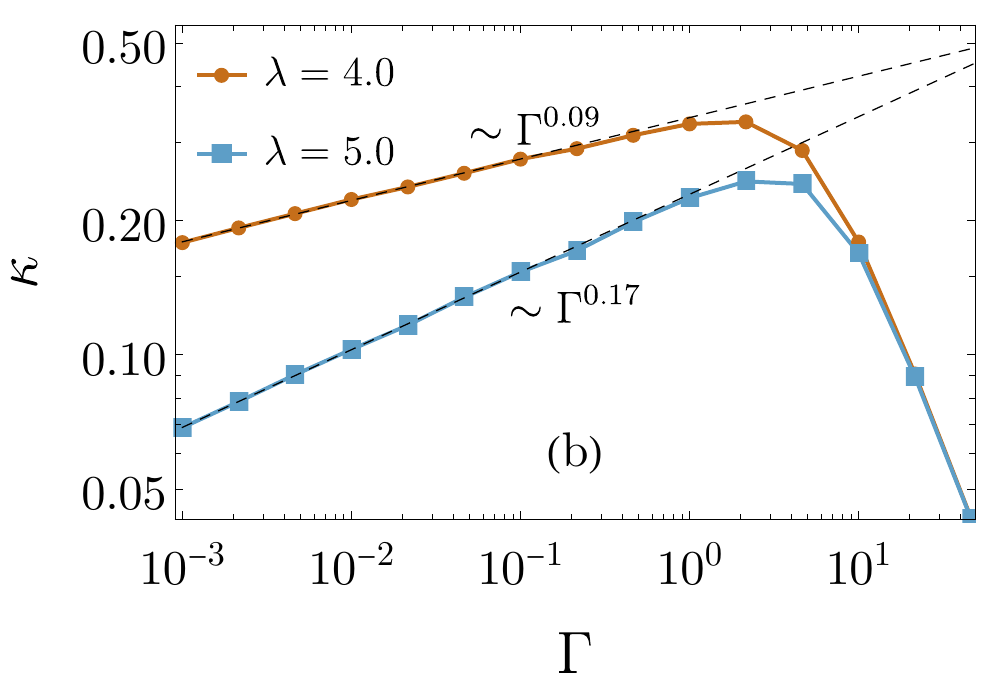}
    \caption{
    (a) $\kappa$~vs.~$\Gamma$ for the Fibonacci model, with $L=987$. 
    Other details are as in Fig.~\ref{fig:fib_deph}. 
    (b) Same, but focusing on the curves for  $\lambda=4$ and $\lambda=5$, for improved visibility.}
    \label{fig:dephasing_enhanced_fib}
\end{figure}

In Figs.~\ref{fig:dephasing_enhanced_aah} and~\ref{fig:dephasing_enhanced_fib} we show the scaling of the conductivity in the AAH and Fibonacci models, for different values of $\lambda$. 
For sufficiently large $\Gamma$, all curves collapse towards the scaling~\eqref{eq:kappa_deph_large_gamma}, regardless of the value of $\lambda$. 
In contrast, when $\Gamma$ is small, different scalings are observed. 
A particularly clear illustration of the change in scalings is the diffusive case of the Fibonacci model (Fig.~\ref{fig:dephasing_enhanced_fib}),  which occurs for $\lambda \approx 3$; the conductivity remains  virtually constant when $\Gamma\to0$, thus recovering the original conductivity of the model without dephasing. 

No we explore the effect of dephasing in the regimes where both models are seen to display subdiffusion; 
in Fig.~\ref{fig:dephasing_enhanced_aah} corresponds to $\lambda > 1$ and in Fig.~\ref{fig:dephasing_enhanced_fib} (a), to $\lambda=4.0$ and $5.0$; the latter are also highlighted separately in Fig.~\ref{fig:dephasing_enhanced_fib} (b), for better visibility. 
These curves represent instances of noise-enhanced transport. That is, where the presence of dephasing actually improves the conductivity. 
As can be seen, this reflects the competition between the scalings \eqref{eq:kappa_deph_small_gamma} and~\eqref{eq:kappa_deph_large_gamma}, for small and large $\Gamma$ respectively. 

\begin{figure}[h!]
    \centering
    \includegraphics[width=0.4\textwidth]{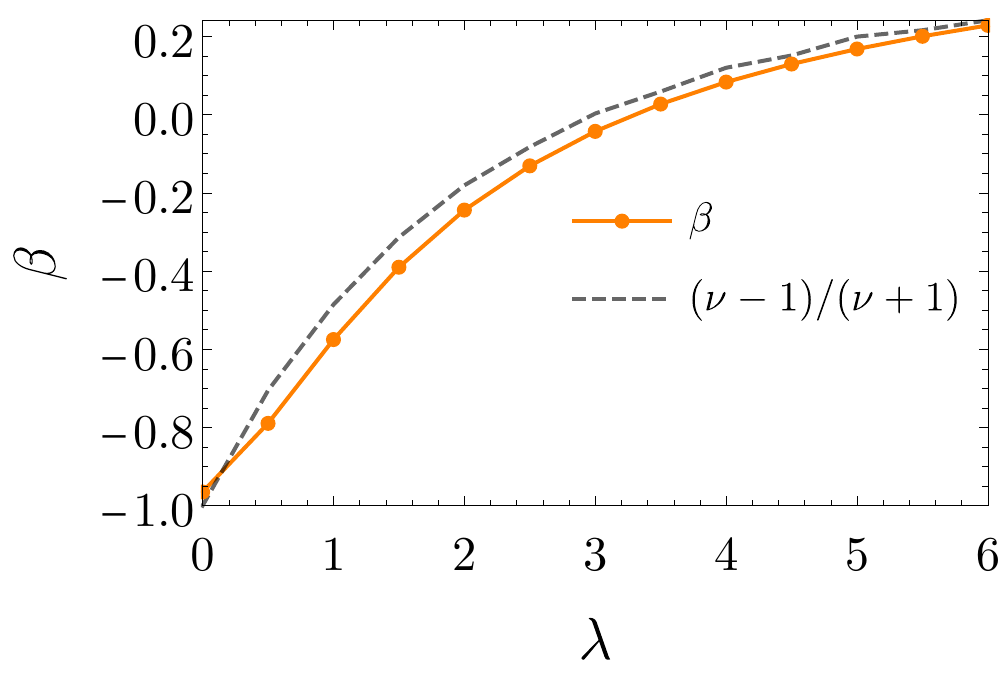}
    \caption[Coefficient $\beta$ fitted from the relation $\kappa \sim \Gamma^\beta$]{Coefficient $\beta$ computed by fitting a curve of the form $\kappa \sim \Gamma^\beta$ in the small $\Gamma$ region. The dashed line shows the value predicted by Eq.~\eqref{eq:kappa_deph_small_gamma}, using the values of $\nu$ shown in Fig.~\ref{fig:no_deph}(d).}
    \label{fig:fib_coef_comparison}
\end{figure}

The small $\Gamma$ behavior predicted by Eq.~\eqref{eq:kappa_deph_small_gamma} is analyzed in Fig.~\ref{fig:fib_coef_comparison} for the Fibonacci model. 
To build this, we focus on the small $\Gamma$ section of all curves in Fig.~\ref{fig:dephasing_enhanced_fib}, and fit a power-law of the form $\kappa \sim \Gamma^\beta$, for some exponent $\beta$. 
This is contrasted with the predictions 
from Eq.~\eqref{eq:kappa_deph_small_gamma}, with $\nu$ determined from Fig.~\ref{fig:no_deph}(d).

{\color{black} We notice that the particular size $L=987$ used in the simulation, which is the larger Fibonacci number we were able to simulate with dephasing, was chosen only for consistency with Sec.~\ref{sec:analysis}. The curves on Figs.~\ref{fig:dephasing_enhanced_aah} and \ref{fig:dephasing_enhanced_fib} are not sensitive to this size being a Fibonacci number.}

\section{\label{sec:discussion}Discussion}

We have undertaken an analysis of the interplay between dephasing and quasiperiodicity in free fermion models. 
Our focus was on boundary driven quantum master equations, which drive the system towards a NESS. 
As we have shown, depending on the model one may obtain a rich variety of transport coefficients which is seen from finite size scaling. 
The AAH model presents clear separations between phases with different behavior; 
conversely, in the Fibonacci model the transport is anomalous and can be tuned continuously by varying the disorder strength. 
In both cases,  when dephasing is present, diffusion emerges. 
Depending on the strength of the quasi-periodic potential, this may give rise to  noise-induced transport, where the dephasing  increases the system's conductivity. 
Our results also show that when the dephasing strength is sufficiently low, the conductivity behaves in a piece-wise fashion as a function of the system size $L$. 
The use of master equations greatly simplify the analysis and is not expected to interfere with the transport coefficients. 

Natural extensions of this analysis include interacting versions of the models~\cite{vznidarivc2018interaction,PhysRevB.103.184205,10.21468/SciPostPhys.6.4.050}, as well as geometries beyond 1D and finite temperatures~\cite{purkayastha2021periodically, PhysRevX.10.031040} and how a combination of these extensions can give rise to further possibilities to exploit dephasing enhanced transport for applications in thermal devices~\cite{Benenti2017a}.

\begin{acknowledgements}
The authors acknowledge A. Purkayastha for fruitful discussions. This work was supported by funding from Science Foundation Ireland and a SFI-Royal Society University Research Fellowship (J. G.). J. G. acknowledges funding from European Research Council Starting Grant ODYSSEY (Grant Agreement No. 758403).
\end{acknowledgements}

\bibliography{library}
\end{document}